\documentclass[11pt,english,epsf]{scrartcl}
\usepackage{lmodern}
\usepackage[T1]{fontenc}
\usepackage[latin9]{inputenc}
\usepackage{geometry}
\usepackage{amsmath}
\usepackage{amssymb}

\newtheorem{theorem}{Theorem}
\newtheorem{lemma}{Lemma}
\newcommand {\dfn} {\stackrel{\Delta} {=}}

\newcommand{\eqa}{\stackrel{\mbox{(a)}}{=}}

\newcommand{\leb}{\stackrel{\mbox{(b)}}{\le}}

\newcommand {\bs} {\mbox{\boldmath $s$}}
\newcommand {\bsigma} {\mbox{\boldmath $\sigma$}}
\newcommand {\btheta} {\mbox{\boldmath $\theta$}}
\newcommand {\bomega} {\mbox{\boldmath $\omega$}}
\newcommand {\bt} {\mbox{\boldmath $t$}}

\newcommand {\bu} {\mbox{\boldmath $u$}}

\newcommand {\bw} {\mbox{\boldmath $w$}}
\newcommand {\bx} {\mbox{\boldmath $x$}}
\newcommand {\by} {\mbox{\boldmath $y$}}
\newcommand {\bz} {\mbox{\boldmath $z$}}

\newcommand {\bE} {\mbox{\boldmath $E$}}

\newcommand {\bX} {\mbox{\boldmath $X$}}

\newcommand {\bZ} {\mbox{\boldmath $Z$}}
\newcommand{\calA}{{\cal A}}

\newcommand{\calC}{{\cal C}}

\newcommand{\calE}{{\cal E}}

\newcommand{\calT}{{\cal T}}

\newcommand{\calX}{{\cal X}}
\newcommand{\calY}{{\cal Y}}
\newcommand{\calZ}{{\cal Z}}
\allowdisplaybreaks

\begin{document}
\thispagestyle{empty}
\title{Universal Decoding Using a Noisy Codebook
}
\author{Neri Merhav
}
\date{}
\maketitle

\begin{center}
The Andrew \& Erna Viterbi Faculty of Electrical Engineering\\
Technion - Israel Institute of Technology \\
Technion City, Haifa 32000, ISRAEL \\
E--mail: {\tt merhav@ee.technion.ac.il}\\
\end{center}
\vspace{1.5\baselineskip}
\setlength{\baselineskip}{1.5\baselineskip}

\begin{center}
{\bf Abstract}
\end{center}
\setlength{\baselineskip}{0.5\baselineskip}
We consider the topic of universal decoding
with a decoder that does not have direct access to the codebook, but only to noisy
versions of the various randomly generated codewords, a problem motivated by
biometrical identification systems. Both the source
that generates the original (clean) codewords, and the channel that corrupts
them in generating the noisy codewords, as well as the main channel for 
communicating the messages, are all modeled by non--unifilar, finite--state
systems (hidden Markov models). As in previous works on universal decoding, 
here too, the average error probability of our proposed universal decoder 
is shown to be as small as that of the optimal maximum likelihood (ML) decoder, 
up to a multiplicative factor
that is a sub--exponential 
function of the block length. It therefore has the same
error exponent, whenever the ML decoder has a positive error exponent.
The universal decoding
metric is based on
Lempel--Ziv (LZ) incremental parsing of each noisy codeword jointly with the given
channel output vector, but this metric is somewhat different from the one
proposed in earlier works on universal decoding for finite--state channels, 
by Ziv (1985) and by Lapidoth and Ziv (1998). The reason for the difference is
that here, unlike in those earlier works, the probability distribution that
governs the (noisy) codewords is, in general, not uniform across its support.
This non--uniformity of the codeword distribution also makes our derivation more
challenging. Another reason for the more challenging analysis is the fact that
the effective induced channel between the noisy codeword of the transmitted
message and the main channel output
is not a finite--state channel in general.\\

\vspace{0.2cm}


\setlength{\baselineskip}{1.5\baselineskip}
\newpage

\section{Introduction}

The topic of universal decoding under channel uncertainty has
received considerable attention in the last four decades. In \cite{Goppa75}
the {\it maximum mutual
information} (MMI) decoder was first proposed
and shown to achieve the capacity for
discrete memoryless channels (DMC's).
Csisz\'ar and K\"orner \cite{CK11} 
showed that the
random coding error exponent of the MMI decoder, associated with a uniform
random coding distribution over a given type class, achieves the same random
coding error exponent as the maximum likelihood (ML) decoder.
Csisz\'ar \cite{Csiszar82} proved that
for any modulo--additive DMC and the uniform
random coding distribution over linear codes, the optimum random coding error
exponent is
universally achieved by a decoder that
minimizes the empirical entropy of the difference between the output sequence
and the input sequence. In \cite{Merhav93}, a parallel result was obtained for
a certain class of memoryless Gaussian channels with slow fading and an unknown
interference signal. 

For channels with memory,
Ziv \cite{Ziv85} considered universal decoding for unknown
unifilar finite--state (FS) channels
with finite input
and output alphabets, i.e., FS channels for which at each time instant, 
the next channel state is given by an unknown deterministic
function of the channel current state, input and output. For ensembles of codes governed by
the uniform distribution over a given permutation--invariant 
set of channel input vectors 
(namely, a type class or the disjoint union of several type classes), 
he proved that a decoder based on the
Lempel--Ziv (LZ) incremental parsing algorithm 
asymptotically achieves the same error exponent as the 
ML decoder. In \cite{LaZ98}, Lapidoth and Ziv proved that the same universal decoder
continues to be universally
asymptotically optimum even for the
broader class of FS channels with stochastic, rather than
deterministic, next--state transitions. They still assumed a random coding
distribution which is uniform over a given permutation--invariant set.
In \cite{FL98}, Feder and Lapidoth have furnished sufficient conditions for
general families of channels with memory
to have universal decoders that asymptotically achieve the random coding error
exponent of ML decoding. In \cite{FM02},
a competitive minimax criterion was proposed, in the quest for a
more general systematic approach to the problem of universal decoding.
Two additional related works on general methodologies for universal
decoding are those of \cite{LF12} and \cite{givenclassofmetrics}.

This paper is a further development on \cite{LaZ98} and \cite{Ziv85}. 
In particular, here we consider universal decoding in a situation
where the decoder that does not have direct access to the codebook of the
encoder, but only to noisy
versions of the various randomly generated codewords, a problem motivated by
applications in biometrical identification systems (see, e.g., \cite[Section
5]{IW10}, \cite{Tuncel09}, \cite{WKGL03}, \cite{WKBL},
and many references therein) or other
applications where storage, or finite--precision limitations do not enable
the decoder to save the exact codewords of all messages, and then they must be
quantized and hence distorted. In our model,
both the source
that generates the original (clean) codewords, and the channel that corrupts
them in the process of generating the noisy codewords, as well as the main channel for
communicating the messages, are all modeled by non--unifilar, FS
systems (hidden Markov models). As in 
the previous above--mentioned works on universal decoding,
here too, the average error probability of our proposed universal decoder
is shown to be as small as that of the optimal maximum likelihood (ML)
decoder, up to a multiplicative factor
that is a sub--exponential
function of the block length, $n$. It therefore has the same
error exponent, whenever the ML decoder has a positive error exponent.
As in \cite{LaZ98} and \cite{Ziv85}, the universal decoding
metric is based on
Lempel--Ziv (LZ) incremental parsing of each noisy codeword jointly with the
given channel output vector, but this metric is somewhat different from that
of \cite{LaZ98} and \cite{Ziv85}.
Specifically, it includes an additional term, which is the logarithm of the
induced probability 
of generating the noisy codeword of the message being tested.
The reason for this difference is
that here, unlike in \cite{LaZ98} and \cite{Ziv85}, the probability
distribution which
governs the (noisy) codewords is, in general, not uniform across its support.
This non--uniformity of the codeword distribution also makes our derivation
quite more challenging. Another factor that makes the analysis here 
more involved is the fact that
the effective induced channel between the noisy codeword of the transmitted
message and the main channel output
is not a FS channel in general.

The outline of the rest of the paper is as follows. In Section 2, we establish
the notation conventions, define the problem formally, and spell out the
assumptions. Section 3 is devoted to the statement of the main result and
a discussion. Finally, in Section 4 the main results is proved.

\section{Notation Conventions, Problem Formulation and Assumptions}

\subsection{Notation Conventions}

Throughout the paper, random variables will be denoted by capital
letters, specific values they may take will be denoted by the
corresponding lower case letters, and their alphabets
will be denoted by calligraphic letters. Random
vectors and their realizations will be denoted,
respectively, by capital letters and the corresponding lower case letters,
both in the bold face font. Their alphabets will be superscripted by their
dimensions. For example, the random vector $\bX=(X_1,\ldots,X_n)$, ($n$ --
positive integer) may take a specific vector value $\bx=(x_1,\ldots,x_n)$
in $\calX^n$, the $n$--th order Cartesian power of $\calX$, which is
the alphabet of each component of this vector.
The probability of an event $\calE$ (with respect to) w.r.t.\ a probability measure $P$ 
will be denoted by $P[\calE]$,
and the expectation
operator w.r.t.\ $P$ will be denoted by
$\bE_P\{\cdot\}$. The subscript will be omitted if the underlying
probability distribution is clear from the context.
Logarithms and exponents will be defined w.r.t.\ the natural basis $e$, unless specified
otherwise. In particular, $\exp_2(t)$ will sometimes be used to denote $2^t$.
The cardinality of a finite set, say, $\calA$, will be denoted by $|\calA|$.

\subsection{Problem Formulation and Assumptions}

Consider a coded communication system, defined as follows. First, a rate--$R$ block code
of length $n$, 
$\{\bx_1,\bx_2,\ldots,\bx_M\}$, $M=e^{nR}$, is
selected at random, where each
$\bx_m\in\calX^n$, $m=1,2,\ldots,M$,
is drawn independently under a distribution $G(\bx)$. A message $m$ is selected
under the uniform distribution over the index set $\{1,2,\ldots,M\}$, and
accordingly, the codeword $\bx_m$ is transmitted over a vector channel
$W(\bz|\bx)$, henceforth referred to as the {\it primary channel} (or the {\it
main channel}),
and the resulting channel output vector, $\bz\in\calZ^n$, is received at the decoder side.
The decoder, however, does not have access to the codebook,
$\{\bx_1,\bx_2,\ldots,\bx_M\}$, used by the encoder, but instead, it has
access to a noisy version of that codebook,
$\calC=\{\by_1,\by_2,\ldots,\by_M\}$, $\by_m\in\calY^n$, $m=1,2,\ldots,M$,
where each $\by_m$ is generated from the corresponding $\bx_m$
by another channel, $V(\by|\bx)$, henceforth referred to as the {\it secondary
channel}. Clearly, this model, which was addressed by Willems {\it et al.} in \cite{WKGL03} and
\cite{WKBL} with application to biometrical identification 
systems (and later, further developed by Tuncel \cite{Tuncel09} and
others), is formally equivalent to the ordinary model of channel random coding, where
the codebook $\calC$ is selected at random, with each member, $\by_m$,
being drawn independently under the random coding distribution,
\begin{equation}
P(\by)=\sum_{\bx\in\calX^n}G(\bx)V(\by|\bx),
\end{equation}
and where upon selecting the index $m$ of the transmitted message, the
corresponding codeword, $\by_m$, is transmitted over the channel
\begin{equation}
\label{echannel}
P(\bz|\by)=\frac{P(\by,\bz)}{P(\by)}=\frac{\sum_{\bx\in\calX^n}G(\bx)V(\by|\bx)W(\bz|\bx)}
{\sum_{\bx\in\calX^n}G(\bx)V(\by|\bx)}.
\end{equation}
From this point onward, the original codebook $\{\bx_1,\bx_2,\ldots,\bx_M\}$
no longer plays a role. Accordingly, we henceforth refer to
$\{P(\by),\by\in\calY^n\}$ as the {\it induced random coding distribution} (or
the {\it effective random coding distribution}), and to
$\{P(\bz|\by)~\by\in\calY^n,~\bz\in\calZ^n\}$ -- as the {\it induced channel} (or
the {\it effective channel}). Clearly, if $G$ is a discrete memoryless source
(DMS) and $V$ is a discrete memoryless channel (DMC), then
$\{P(\by),\by\in\calY^n\}$ is a DMS as well. If, in addition, $W$ is also a DMC,
then so is the channel $\{P(\bz|\by)~\by\in\calY^n,~\bz\in\calZ^n\}$.
In this case, the
capacity of the system is simply the mutual information, $I(Y;Z)$, pertaining to
the single--letter marginal $\{P(y,z),~y\in\calY,~z\in\calZ\}$, see \cite{WKGL03}, \cite{WKBL}.
It should be noted, however, that unlike the traditional model of random
coding for channels, where random coding is a technical concept that merely serves the
purpose of proving existence of good codes, here, when it comes to
biometrical systems applications, the randomness of the code is
part of the model setting. As a consequence, both $G$ and $V$,
and hence also the induced
random coding distribution, $\{P(\by),\by\in\calY^n\}$, are dictated to us,
and are not subjected to our control.\footnote{For this reason, the capacity is
simply given by $I(Y;Z)$, without maximizing over the distribution of $Y$.}

As in \cite{WKGL03}, \cite{WKBL}, here too, it is assumed that all three 
alphabets, $\calX$, $\calY$,
and $\calZ$, are finite.
In this paper, however, we go considerably beyond the realm of
memoryless systems, and allow $G$, $V$ and $W$ to be
all non-unifilar, FS systems (hidden
Markov models), as follows. 
The distribution $G$ assumes the form
\begin{equation}
\label{source}
G(\bx)=\sum_{\bomega}\prod_{i=1}^n G(x_i,\omega_i|\omega_{i-1}),
\end{equation}
where $\bx$ is as before, $\bomega=(\omega_1,\ldots,\omega_n)$ is the source state vector,
whose components take on values in a finite set $\Omega$, and the initial
state, $\omega_0$ is assumed fixed.
The primary channel, $W$, is modeled as
\begin{equation}
\label{wchannel}
W(\bz|\bx)=\sum_{\bsigma}\prod_{i=1}^n W(z_i,\sigma_i|x_i,\sigma_{i-1}),
\end{equation}
where $\bsigma=(\sigma_1,\ldots,\sigma_n)$ is the channel state vector, whose
components take on values in a finite set
$\Sigma$ and the initial state, $\sigma_0$, is fixed.
Likewise, the secondary channel, $V$, is given by
\begin{equation}
\label{vchannel}
V(\by|\bx)=\sum_{\btheta}\prod_{i=1}^n V(y_i,\theta_i|x_i,\theta_{i-1}),
\end{equation}
where $\btheta=(\theta_1,\ldots,\theta_n)$ is the state vector whose components take on
values in a finite set
$\Theta$ and there is fixed initial state, $\theta_0$.

We consider the problem of universal decoding for the effective channel
$P(\bz|\by)$ induced by the source (\ref{source}), the main channel
(\ref{wchannel}) and the secondary channel (\ref{vchannel}), according to
(\ref{echannel}). We will assume that $G$, $V$ and $W$ 
are not known to the decoder, and hence nor is
the effective channel $\{P(\bz|\by)~\by\in\calY^n,\bz\in\calZ^n\}$.
Nonetheless, the
effective random
coding distribution, $\{P(\by),~\by\in\calY^n\}$, will assumed known to
the decoder. The rationale behind the latter assumption
stems from the fact
that the decoder knows the codebook, $\calC=\{\by_1,\ldots,\by_M\}$, and so,
it has access to
an exponential amount of data from which the parameters of this
distribution can be estimated very accurately. In particular, note that
$P(\by)$ has a hidden Markov structure,
\begin{eqnarray}
\label{hmm1}
P(\by)&=&\sum_{\bx} G(\bx)V(\by|\bx)\nonumber\\
&=&\sum_{\btheta,\bomega,\bx}\prod_{i=1}^n
G(x_i,\omega_i|\omega_{i-1})V(y_i,\theta_i|x_i,\theta_{i-1})\nonumber\\
&=&\sum_{\btheta,\bomega}\prod_{i=1}^n\left[\sum_xG(x,\omega_i|\omega_{i-1})
V(y_i,\theta_i|x,\theta_{i-1})\right]\nonumber\\
&=&\sum_{\btheta,\bomega}\prod_{i=1}^n\pi(y_i,\theta_i,\omega_i|\theta_{i-1},\omega_{i-1}),
\end{eqnarray}
where in the last passage, we have defined the parameters
$\pi(y,\theta,\omega|\theta',\omega')\dfn\sum_x
G(x,\omega|\omega')V(y,\theta|x,\theta')$. These parameters
can be estimated using well known estimation methods for hidden Markov
models.\footnote{
The ML estimator for the parameters of a hidden Markov model, 
is known to be strongly
consistent \cite{BP66}, \cite{Petrie69}. More practically, one may use the
iterative Baum algorithm, which is an instance of the EM
algorithm \cite{DLR77} (see also the tutorials 
\cite{EM02}, \cite{Rabiner89} and references
therein).} It will be assumed\footnote{Note that this assumption concerns
$G$ and $V$ only, it has nothing to do with the primary channel $W$. If the parameters
$\{\pi(y,\theta,\omega|\theta',\omega')\}$ are estimated using the ML
estimator (referring to footnote 2), 
then eq.\ (\ref{positivity}) can be imposed as a constraint on the
estimator.}
that
\begin{equation}
\label{positivity}
\pi(y,\theta,\omega|\theta',\omega') > 0
\end{equation}
for all $(\omega,\omega',\theta,\theta',y)\in\Omega^2\times\Theta^2\times\calY$, and we denote
$\pi_{\min}\dfn
\min_{\omega,\omega',\theta,\theta',y}\pi(y,\theta,\omega|\theta',\omega')$.

Like in previous works on universal decoding, our objective is
to devise a universal decoding metric whose average error probability
is of the same exponential order as that of the ML decoder.
As described in the Introduction,
the problem of universal decoding for FS channels was considered first in
\cite{Ziv85}, where it was assumed that the next--state transitions are given by a
deterministic function of the current state, the current input and the
current output. In \cite{LaZ98}, the framework was extended to handle
general FS channels, where the state transitions were allowed to be stochastic
(as in eqs.\ (\ref{wchannel}) and (\ref{vchannel}) above). Also, in both
\cite{LaZ98} and
\cite{Ziv85}, the random coding distribution was assumed uniform across a
given permutation--invariant set.\footnote{A permutation--invariant set is a
set that is closed under permutations, in other words, a set that can be
represented by the disjoint union of type classes.} 
Here the situation
is different from both \cite{LaZ98} and \cite{Ziv85} because of two reasons.
\begin{enumerate}
\item The effective random coding distribution $\{P(\by),~\by\in\calY^n\}$ 
is not uniform over a
permutation--invariant set, in general.
\item The effective channel $\{P(\bz|\by),~\by\in\calY^n,~\bz\in\calZ^n\}$ 
is not a FS channel, in general.
\end{enumerate}
These differences are important, because in \cite{LaZ98} and \cite{Ziv85},
both assumptions were used rather heavily.

For a given noisy code $\calC$ and a given channel output vector $\bz$,
let us define (similarly as in \cite{FL98} and \cite{LaZ98}) the ranking of
the members of $\calY^n$, according to descending likelihood values, 
i.e., $P(\bz|\by[1])\ge P(\bz|\by[2])\ge \ldots$, and let us denote by
$M_{\mbox{\tiny o}}(\by,\bz)$ the ranking of $\by$ given $\bz$.
For a given $\bz$, the ranking function $M_{\mbox{\tiny o}}(\by,\bz)$ is
therefore a
one--to--one mapping from $\calY^n$ to the set $\{1,2,\ldots,|\calY|^n\}$
with the property that $P(\bz|\by')> P(\bz|\by)$ implies $M_{\mbox{\tiny
o}}(\by',\bz)< M_{\mbox{\tiny o}}(\by,\bz)$.
The probability of error associated with the
ML decoder for the given code $\calC$ and the effective channel,
$\{P(\bz|\by),~\by\in\calY^n,~\bz\in\calZ^n\}$, is given by
\begin{equation}
\mbox{P}_{\mbox{\tiny e,o}}(\calC)=\frac{1}{M}\sum_{m=1}^M
P\left[\bigcup_{m'\ne m}\left\{M_{\mbox{\tiny o}}(\by_{m'},\bZ)\le M_{\mbox{\tiny
o}}(\by_m,\bZ)\right\}
\bigg|\mbox{message $m$ was sent}\right],
\end{equation}
where the event $M_{\mbox{\tiny o}}(\by_{m'},\bZ)= M_{\mbox{\tiny
o}}(\by_m,\bZ)$ accounts for the case where $\by_{m'}=\by_m$ (which is
possible since the members of $\calC$ are chosen independently at random).
The average probability of error w.r.t.\ the randomness of $\calC$, is
then
\begin{eqnarray}
\overline{\mbox{P}}_{\mbox{\tiny e,o}}&=&\bE\left\{\mbox{P}_{\mbox{\tiny
e,o}}(\calC)\right\}\\
&=&1-\sum_{\by,\bz}P(\by,\bz)\left(1-P[\calE_{\mbox{\tiny o}}(\by,\bz)]\right)^{e^{nR}-1},
\end{eqnarray}
where
\begin{equation}
\calE_{\mbox{\tiny o}}(\by,\bz)\dfn \{\by':~M_{\mbox{\tiny o}}(\by',\bz)\le M_{\mbox{\tiny
o}}(\by,\bz)\}.
\end{equation}
As in \cite{FL98} and \cite{LaZ98}, for later use, we define the function
\begin{equation}
\label{f-def}
f(t)\dfn 1-(1-t)^{e^{nR}-1},~~~~t\in[0,1],
\end{equation}
and so,
\begin{equation}
\overline{\mbox{P}}_{\mbox{\tiny
e,o}}=\sum_{\by,\bz}P(\by,\bz)f\left(P[\calE_{\mbox{\tiny
o}}(\by,\bz)]\right).
\end{equation}
By the same token, for an arbitrary decoding metric $u(\by,\bz)$, we define
a ranking function $M_{\mbox{\tiny u}}(\by,\bz)$, as any one--to--one mapping
$\calY^n:\to\{1,2,\ldots,|\calY|^n\}$ given $\bz$, such that
$\bu(\by',\bz) < \bu(\by,\bz)$ implies $M_{\mbox{\tiny u}}(\by',\bz)<
M_{\mbox{\tiny u}}(\by,\bz)$. Accordingly, the average error probability
associated with $u(\cdot,\cdot)$, is given by
\begin{equation}
\overline{\mbox{P}}_{\mbox{\tiny
e,u}}=\sum_{\by,\bz}P(\by,\bz)f\left(P[\calE_{\mbox{\tiny
u}}(\by,\bz)]\right),
\end{equation}
where
\begin{equation}
\calE_{\mbox{\tiny u}}(\by,\bz)\dfn \{\by':~M_{\mbox{\tiny u}}(\by',\bz)\le
M_{\mbox{\tiny
u}}(\by,\bz)\}.
\end{equation}
We are interested in a universal metric $u(\cdot,\cdot)$, that is independent
of the unknown effective channel (but possibly dependent on the
effective random coding distribution), such that $\overline{\mbox{P}}_{\mbox{\tiny
e,u}}$ would not exceed $\overline{\mbox{P}}_{\mbox{\tiny e,o}}$ by more than
a sub--exponential function of $n$, i.e.,
\begin{equation}
\label{ao}
\overline{\mbox{P}}_{\mbox{\tiny e,u}}\le e^{n\epsilon(n)}
\overline{\mbox{P}}_{\mbox{\tiny e,o}},
\end{equation}
where $\epsilon(n)\to 0$ as $n\to\infty$.

\section{Main Result}

Given two sequences, $\by$ and $\bz$, both of length $n$, consider 
the joint incremental parsing \cite{ZL78} of the
sequence of pairs
$$(y_1,z_1),(y_2,z_2),\ldots,(y_n,z_n)$$
into $c$ distinct
phrases. Specifically, denoting $w_i=(y_i,z_i)$, $i=1,2,\ldots,n$, we parse
$\bw=(w_1,\ldots,w_n)$, sequentially into the distinct\footnote{
To be more precise, the phrases are all distinct with 
the possible exception of the last phrase,
which might be incomplete.} phrases,
$w_1^{n_1},w_{n_1+1}^{n_2},\ldots,w_{n_{c-1}+1}^n$, where $n_i+1$ is the
starting point of the $i$--th phrase, $i=1,2,\ldots,c$ ($n_0=0$). According to
the incremental parsing procedure of the LZ algorthm, 
each phrase $w_{n_i+1}^{n_{i+1}}$ is the
shortest string that has not been encountered before as a parsed phrase, which
means that its prefix, $w_{n_i+1}^{n_{i+1}-1}$, is identical to an earlier
phrase, $w_{n_j+1}^{n_{j+1}}$, $j<i$.
Let $c\equiv c(\by,\bz)$ denote the number of distinct phrases.
For example,\footnote{The same example appears in \cite{Ziv85}.} if
\begin{eqnarray}
\by&=&0~|~1~|~0~0~|~0~1|\nonumber\\
\bz&=&0~|~1~|~0~1~|~0~1|\nonumber
\end{eqnarray}
then $c(\by,\bz)=4$.
Let $c(\bz)$ denote the resulting number of distinct phrases
of $\bz$, and let $\bz(\ell)$ denote the $\ell$th distinct $\bz$--phrase,
$\ell=1,2,...,c(\bz)$. In the above example, $c(\bz)=3$. Denote by
$c_\ell(\by|\bz)$ the number of occurrences of $\bz(\ell)$ in the
parsing of $\bz$, or equivalently, the number of distinct $\by$-phrases
that jointly appear with $\bz(\ell)$. Clearly, $\sum_{\ell=1}^{c(\bz)} c_\ell(\by|\bz)=
c(\by,\bz)$. In the above example, $\bz(1)=0$, $\bz(2)=1$, $\bz(3)=01$,
$c_1(\by|\bz)=c_2(\by|\bz)=1$, and $c_3(\by|\bz)=2$. We next define 
our universal decoding metric as
\begin{equation}
\label{umetric}
u(\by,\bz)\dfn\log P(\by)+\sum_{\ell=1}^{c(\bz)}c_\ell(\by|\bz)\log c_\ell(\by|\bz),
\end{equation}
which in turn, defines the decoder
\begin{equation}
\label{udecoder}
\hat{m}_{\mbox{\tiny u}}=\mbox{arg min}_m u(\by_m,\bz),
\end{equation}
where ties broken according to an arbitrary ranking function $M_{\mbox{\tiny
u}}(\cdot,\bz)$ associated with (\ref{umetric}).

We are now ready to state our main result, whose proof appears in Section 4.

\begin{theorem}
Under the assumptions of Subsection 2.2, the universal decoder
(\ref{udecoder}) satisfies eq.\ (\ref{ao}) where
$\epsilon(n)=O((\log\log n)/\log n)$, with a leading term\footnote{The
sequence $\epsilon(n)$ depends also on other parameters of the problem, like
$|\Theta|$, $|\Omega|$, $|\Sigma|$, and $\pi_{\min}$, but these parameters
appear in negligible terms of $\epsilon(n)$, that decay faster than $(\log\log n)/\log n$.}
that is linear in 
$\log|\calY\times\calZ|$.
\end{theorem}

It should be noticed that the universal decoding metric (\ref{umetric}) is
different from the one in \cite{LaZ98} and \cite{Ziv85}, because it includes
the term $\log P(\by)$ in addition to the LZ conditional
compressibility term, $\sum_{\ell=1}^{c(\bz)}c_\ell(\by|\bz)\log
c_\ell(\by|\bz)$ (see also \cite{givenclassofmetrics}). 
The reason for this difference is that the effective random
coding distribution, $\{P(\by),~\by\in\calY^n\}$, is not necessarily uniform
over its support, in contrast to the assumption in both \cite{LaZ98} and
\cite{Ziv85}. In a way, the decoder (\ref{udecoder}) can be seen as an
extension of the MMI decoder, which is the well known
universal decoder for DMCs \cite{CK11}. To see this, observe that
(\ref{udecoder}) can be rewritten as
\begin{equation}
\label{alternative}
\hat{m}_{\mbox{\tiny u}}=\mbox{arg max}_m
\left\{\frac{1}{n}\log\left[\frac{1}{P(\by_m)}\right]-\frac{1}{n}\sum_{\ell=1}^{c(\bz)}c_\ell(\by_m|\bz)\log
c_\ell(\by_m|\bz)\right\},
\end{equation}
where the term $\frac{1}{n}\log[1/P(\by_m)]$ plays a role like the
empirical entropy associated with $\by_m$ and the term
$\frac{1}{n}\sum_{\ell=1}^{c(\bz)}c_\ell(\by_m|\bz)\log c_\ell(\by_m|\bz)$ is
parallel to the conditional empirical entropy of $\by_m$ given $\bz$. Thus,
the difference is analogous to a certain notion of a generalized empirical mutual information.
But having said that, we should add a digression that, when confining the discussion
to the memoryless case, the
first term in (\ref{alternative}) gives the empirical entropy of $\by_m$ 
only in the case where $\{P(\by)\}$ is uniform across a single type class. If
instead, it is a product distribution, then the MMI metric should be
supplemented with a divergence term between the empirical distribution and the
true distribution.\footnote{In this context, the author has some doubts
concerning the asymptotic optimality of the MMI decoder used in \cite{DD11}.}

The proof of Theorem 1 contains essentially similar ingredients to those in
\cite{LaZ98}. There are, however, a 
few differences that should be pointed out.
In the previous
paragraph, we mentioned that here, as opposed to those papers, the random
coding distribution is not uniform in general.
This difference is also responsible for the fact that 
there are a few non--trivial issues in the extension of the derivations of
\cite{LaZ98} and \cite{Ziv85} to our setting, as in those two earlier papers,
the uniformity of the random coding distribution (across its support), was
used quite heavily. In particular, the pairwise error probability,
$P[\calE_{\mbox{\tiny o}}(\by,\bz)]$, which plays a central role in the
analysis in \cite{LaZ98} and \cite{Ziv85}, is simply proportional to the
cardinality of $\calE_{\mbox{\tiny o}}(\by,\bz)$, namely to $M_{\mbox{\tiny
o}}(\by,\bz)$, which in turn, can be evaluated 
using combinatorial considerations. Here, on the other hand, the members of
$\calE_{\mbox{\tiny o}}(\by,\bz)$ have to be weighed by their various
probabilities, $\{P(\by^\prime),~\by^\prime\in\calE_{\mbox{\tiny o}}(\by,\bz)\}$.
In particular, in an important technical lemma of \cite{LaZ98} (Lemma 2
therein), the last
step of the proof is relatively easy, because thanks to the
uniformity assumption therein, it is associated with
the calculation of the quantity, $\sum_{\by}1/M_{\mbox{\tiny o}}(\by,\bz)$
(in our notation),
which is nothing but the harmonic series, $\sum_{i=1}^{N} 1/i\le\ln N +1$ ($N$
-- positive integer), as $M_{\mbox{\tiny
o}}(\by,\bz)$ is defined as a ranking function (see, in particular, the last step in the chain of
inequalities at the end of page 1751 in \cite{LaZ98}). For the non--uniform input
considered here, the relevant extension of the above mentioned expression
turns out to be $\sum_{\by}P(\by)/P[\calE_{\mbox{\tiny o}}(\by,\bz)]$, which
is not as straightforward to bound in a useful manner. Fortunately enough,
as is shown in Lemma 1 below, this can nevertheless still be done, and in a
quite general manner, that is almost 
completely unrelated to the hidden Markov structure of the
model. Another source for some technical
challenges is the fact that the induced channel,
$\{P(\bz|\by)\}$, is not a FS channel, in general. This calls for separate
treatment of the numerator and the denominator of
$P(\bz|\by)=P(\by,\bz)/P(\by)$ (which both obey a hidden Markov model),
that in turn, may be dominated by two different sequences of states.
Nonetheless, these difficulties can also be circumvented, as
will be seen in Section 4. 

\section{Proof of Theorem 1}

The idea of the proof is to lower bound $\bar{\mbox{P}}_{\mbox{\tiny e,o}}$
and to upper bound
$\bar{\mbox{P}}_{\mbox{\tiny e,u}}$ by two expressions which are identical up
to a multiplicative factor of $e^{n\epsilon(n)}$. We begin with the upper
bound to $\bar{\mbox{P}}_{\mbox{\tiny e,u}}$.

Let us denote 
\begin{equation}
v(\by,\bz)\dfn \sum_{\ell=1}^{c(\bz)}c_\ell(\by|\bz)\log c_\ell(\by|\bz),
\end{equation}
so that $u(\by,\bz)=\log P(\by)+v(\by,\bz)$.
We will use the fact that $v(\by,\bz)$ is almost large enough to serve as a
legitimate length function for lossless compression of $\by$ given $\bz$,
where $\bz$
serves as side information available to both the encoder and the decoder.
In particular, in
the proof of Lemma 2 in \cite[p.\ 460]{Ziv85}, Ziv describes a lossless
compression scheme with side information, whose length function, $L(\by|\bz)$,
satisfies
\begin{equation}
L(\by|\bz)\le v(\by,\bz)+n\epsilon_1(n),
\end{equation}
with 
\begin{equation}
\epsilon_1(n)=O\left(\frac{\log\log n}{\log n}\right),
\end{equation}
whose leading term is linear in $\log|\calY\times\calZ|$.
Now, let us define 
\begin{equation}
\bar{\mbox{P}}_{\mbox{\tiny e,u}}(\by,\bz)=f(P[\calE_{\mbox{\tiny
u}}(\by,\bz)]),
\end{equation}
where $f(\cdot)$ is defined as in (\ref{f-def}). Now,
\begin{eqnarray}
\label{peu}
P[\calE_{\mbox{\tiny u}}(\by,\bz)]&=&
\sum_{\{\by^\prime:~M_{\mbox{\tiny u}}(\by^\prime,\bz)\le M_{\mbox{\tiny
u}}(\by,\bz)\}}P(\by^\prime)\nonumber\\
&\le& \sum_{\{\by^\prime:~P(\by^\prime)\exp_2[
v(\by^\prime,\bz)]\le P(\by)\exp_2[
v(\by,\bz)]\}}P(\by^\prime)\nonumber\\
&\le& \sum_{\{\by^\prime:~P(\by^\prime)\exp_2[
v(\by^\prime,\bz)]\le P(\by)\exp_2[
v(\by,\bz)]\}}P(\by)\exp_2[v(\by,\bz)-
v(\by^\prime,\bz)]\nonumber\\
&\le& P(\by)\exp_2[
v(\by,\bz)]\sum_{\by^\prime\in\calY^n}\exp_2[
-v(\by^\prime,\bz)]\nonumber\\
&\le& 2^{n\epsilon_1(n)}P(\by)\exp_2[
v(\by,\bz)]\sum_{\by^\prime\in\calY^n}2^{-L(\by^\prime|\bz)}\nonumber\\
&\le& e^{n\epsilon_1(n)}P(\by)\cdot\exp_2[
v(\by,\bz)]\nonumber\\
&=&e^{n\epsilon_1(n)}\cdot\exp_2[u(\by,\bz)],
\end{eqnarray}
where the in the second to the last step, we have used Kraft's inequality and we bounded
$2^{n\epsilon_1(n)}$ by $e^{n\epsilon_1(n)}$, simply for convenience in later steps of
the proof. It now follows from (\ref{peu}) and the monotonicity of $f$ that
\begin{equation}
\bar{\mbox{P}}_{\mbox{\tiny e,u}}(\by,\bz)\le
f\left(e^{n\epsilon_1(n)}\cdot\exp_2[u(\by,\bz)]\right).
\end{equation}
For later use, we also have
\begin{eqnarray}
\bar{\mbox{P}}_{\mbox{\tiny e,u}}(\bz)&\dfn&
\sum_{\by\in\calY^n}P(\by|\bz)\bar{\mbox{P}}_{\mbox{\tiny e,u}}(\by,\bz)\\
&\le& \sum_{\by\in\calY^n}P(\by|\bz)
f\left(e^{n\epsilon_1(n)}\cdot\exp_2[u(\by,\bz)]\right).
\end{eqnarray}

We next move on to derive a matching lower bound to $\bar{\mbox{P}}_{\mbox{\tiny e,o}}$.
Similarly, as in \cite{LaZ98}, we will need to refer to an auxiliary threshold
decoder (in the terminology of \cite{LaZ98}), 
which is a slightly more conservative version of the ML decoder.
Specifically, for a given threshold parameter, $\alpha> 1$,
this decoder outputs the message $m$ with the property that $P(\bz|\by_m)> \alpha\cdot
P(\bz|\by_{m^\prime})$ for all $m^\prime\ne m$, and declares an error if no such
$m$ exists. Accordingly, let $\bar{\mbox{P}}_{\mbox{\tiny e,t}}(\by,\bz)$
denote the conditional average error probability of the threshold decoder, given
$(\by,\bz)$, i.e.,
\begin{equation}
\bar{\mbox{P}}_{\mbox{\tiny e,t}}(\by,\bz)=f(P[\calE_{\mbox{\tiny
t}}(\by,\bz)]),
\end{equation}
where
\begin{equation}
\calE_{\mbox{\tiny
t}}(\by,\bz)=\{\by^\prime:~P(\bz|\by^\prime)\ge\alpha^{-1}P(\bz|\by)\}.
\end{equation}
As in Lemma 2 of \cite{LaZ98}, here too, the next lemma (proved in the
appendix) asserts that
the performance of the threshold decoder cannot be much worse than that of the
ML decoder, provided that $\alpha$ is not too large. In particular, if
$\alpha=\alpha_n$ grows subexponetially with $n$, then the threshold decoder
has the same error exponent as that of the ML decoder.

\begin{lemma}
\label{threshold}
Define
\begin{eqnarray}
\bar{\mbox{P}}_{\mbox{\tiny e,t}}(\bz)=\sum_{\by\in\calY^n}P(\by|\bz)
f(P[\calE_{\mbox{\tiny
t}}(\by,\bz)])\\
\bar{\mbox{P}}_{\mbox{\tiny e,o}}(\bz)=\sum_{\by\in\calY^n}P(\by|\bz)
f(P[\calE_{\mbox{\tiny
o}}(\by,\bz)]).
\end{eqnarray}
Then, under the positivity assumption
(\ref{positivity}), 
\begin{equation}
\bar{\mbox{P}}_{\mbox{\tiny e,t}}(\bz)\le
\left\{\alpha\left[n\ln\left(\frac{1}{\pi_{\min}
\cdot|\Theta|\cdot|\Omega|}\right)+1\right]+1\right\}
\cdot\bar{\mbox{P}}_{\mbox{\tiny e,o}}(\bz)
\end{equation}
for every $\bz\in\calZ^n$.
\end{lemma}
It should be noted that assumption (\ref{positivity}) is essentially not needed for
the above Lemma. What is really needed is that the smallest
$P(\by)$, across all $\by\in\calY^n$ with $P(\by)>0$, 
would not decay faster than exponentially
with $n$. But owing to (\ref{hmm1}), one can easily see that $P(\by)\ge
\pi_+^n$, where $\pi_+$ is the smallest positive
$\pi(y,\theta,\omega|\theta^\prime,\omega^\prime)$. We are using
(\ref{positivity}) nonetheless, because we make this assumption anyway (as it
is needed elsewhere), and 
then the upper bound given by the
lemma is slightly tighter.

On the basis of Lemma \ref{threshold}, any lower bound on
$\bar{\mbox{P}}_{\mbox{\tiny e,t}}$ in terms of 
$\bar{\mbox{P}}_{\mbox{\tiny e,u}}$, would immediately yield a lower bound
$\bar{\mbox{P}}_{\mbox{\tiny e,o}}$ in terms of 
$\bar{\mbox{P}}_{\mbox{\tiny e,u}}$, as desired. Accordingly, the next step would be to lower
bound $\bar{\mbox{P}}_{\mbox{\tiny e,t}}$.
This in turn will be done by lower bounding
$P[\calE_{\mbox{\tiny t}}(\by,\bz)]$ (for a certain choice of the
threshold $\alpha$, to be defined) in terms of
$P[\calE_1(\by,\bz)]$, for a certain $\calE_1(\by,\bz)\subseteq
\calE_{\mbox{\tiny t}}(\by,\bz)$ to be specified shortly.

First observe that, similarly as in
eq.\ (\ref{hmm1}),
\begin{eqnarray}
P(\by,\bz)&=&\sum_{\btheta,\bsigma,\bomega,\bx}
\prod_{i=1}^n[G(x_i,\omega_i|\omega_{i-1})
V(y_i,\theta_i|x_i,\theta_{i-1})W(z_i,\sigma_i|x_i,\sigma_{i-1})]\\
&=&\sum_{\btheta,\bsigma,\bomega}\prod_{i=1}^n\sum_x[G(x,\omega_i|\omega_{i-1})
V(y_i,\theta_i|x,\theta_{i-1})W(z_i,\sigma_i|x,\sigma_{i-1})]\\
&=&\sum_{\btheta,\bsigma,\bomega}\prod_{i=1}^n\Pi(y_i,z_i,\theta_i,\sigma_i,\omega_i
|\theta_{i-1},\sigma_{i-1},\omega_{i-1})
\end{eqnarray}
where we have defined
$\Pi(y,z,\theta,\sigma,\omega|\theta',\sigma',\omega')=\sum_xG(x,\omega|\omega')
V(y,\theta|x,\theta')W(z,\sigma|x,\sigma')$.
We will henceforth use the following notation for two positive integers $i$
and $j$, where $j > i$:
\begin{eqnarray}
& &\Pi(y_i^j,z_i^j,\theta_j,\sigma_j,\omega_j|
\theta_{i-1},\sigma_{i-1},\omega_{i-1})\nonumber\\
&=&\sum_{\theta_i^{j-1}}\sum_{\sigma_i^{j-1}}\sum_{\omega_i^{j-1}}
\prod_{k=i}^j\Pi(y_k,z_k,\theta_k,\sigma_k,\omega_k|\theta_{k-1},\sigma_{k-1},\omega_{k-1})
\end{eqnarray}
and
\begin{equation}
\pi(y_i^j,\theta_j,\omega_j|\theta_{i-1},\omega_{i-1})=
\sum_{\theta_i^{j-1}}\sum_{\omega_i^{j-1}}
\prod_{k=i}^j\pi(y_k,\theta_k,\omega_k|\theta_{k-1},\omega_{k-1}).
\end{equation}
Next, define
\begin{eqnarray}
\bt&\dfn&\{(\theta_i,\sigma_i,\omega_i):~i=n_0,n_1,\ldots,n_{c-1}\},\\
\bs&\dfn&\{(\theta_i,\omega_i):~i=n_0,n_1,\ldots,n_{c-1}\},
\end{eqnarray}
where $\{n_i\}$ are phrase boundaries, as defined at the beginning of Section
3, for a given $(\by,\bz)$. Now, for the same $(\by,\bz)$, let
\begin{eqnarray}
\hat{\bt}&=&\mbox{arg max}_{\bt} P(\by,\bz,\bt)=\mbox{arg max}_{\bt}\prod_{i=0}^{c-1}
\Pi(y_{n_i+1}^{n_{i+1}},z_{n_i+1}^{n_{i+1}},\theta_{n_{i+1}},\sigma_{n_{i+1}},\omega_{n_{i+1}}
|\theta_{n_i},\sigma_{n_i},\omega_{n_i})\\
\tilde{\bs}&=&\mbox{arg max}_{\bs} P(\by,\bs)=\mbox{arg max}_{\bs}\prod_{i=0}^{c-1}
\pi(y_{n_i+1}^{n_{i+1}},\theta_{n_{i+1}},\omega_{n_{i+1}}
|\theta_{n_i},\omega_{n_i}).
\end{eqnarray}
We denote the components of $\hat{\bt}$ and $\tilde{\bs}$ by
$\{(\hat{\theta}_{n_i},\hat{\sigma}_{n_i},\hat{\omega}_{n_i})\}$ and
$\{(\tilde{\theta}_{n_i},\tilde{\omega}_{n_i})\}$, respectively.
Denoting $K=|\Theta\times\Sigma\times\Omega|$, it is obvious that
$P(\by,\bz,\hat{\bt})\ge K^{-c}P(\by,\bz)$, and a similar relation holds
between $P(\by,\tilde{\bs})$ and $P(\by)$.
For the given pair $(\by,\bz)$, let 
\begin{equation}
\calE_1(\by,\bz)\dfn\left\{\by':~P(\by',\bz,\hat{\bt})=
P(\by,\bz,\hat{\bt}),~P(\by',\tilde{\bs})=P(\by,\tilde{\bs})\right\}.
\end{equation}
Owing to assumption (\ref{positivity}), it is shown in the appendix 
(similarly as in \cite[eq.\ (A.7)]{ZM92}) that 
\begin{equation}
\label{zm92}
P(\by')\le
P(\by',\tilde{\bs})\cdot\left(\frac{|\Theta\times\Omega|}{\pi_{\min}^2}\right)^c
\le P(\by',\tilde{\bs})\cdot\left(\frac{K}{\pi_{\min}^2}\right)^c, 
\end{equation}
and so, for $\by^\prime\in\calE_1(\by,\bz)$,
the chain of inequalities,
\begin{eqnarray}
\left(\frac{K}{\pi_{\min}^2}\right)^c\cdot P(\bz|\by')&=&
\left(\frac{K}{\pi_{\min}^2}\right)^c\cdot \frac{P(\by',\bz)}{P(\by')}\\
&\ge&\left(\frac{K}{\pi_{\min}^2}\right)^c\frac{P(\by',\bz,\hat{\bt})}
{(K/\pi_{\min}^2)^cP(\by',\tilde{\bs})}\\
&=&\frac{P(\by',\bz,\hat{\bt})}{P(\by',\tilde{\bs})}\\
&=&\frac{P(\by,\bz,\hat{\bt})}{P(\by,\tilde{\bs})}\\
&\ge&K^{-c}\frac{P(\by,\bz)}{P(\by)}\\
&=&K^{-c}P(\bz|\by),
\end{eqnarray}
implies that
\begin{eqnarray}
\calE_1(\by,\bz)&\subseteq& 
\{\by':~P(\bz|\by')\ge (K/\pi_{\min})^{-2c}P(\bz|\by)\}\\
&\subseteq&\{\by':~P(\bz|\by')\ge (K/\pi_{\min})^{-2\bar{c}_n}P(\bz|\by)\}\\
&=&\calE_{\mbox{\tiny t}}(\by,\bz)~~~~~\mbox{with the
choice}~\alpha=(K/\pi_{\min})^{2\bar{c}_n}
\end{eqnarray}
where 
\begin{equation}
\bar{c}_n\dfn\frac{n\log|\calY\times\calZ|}{(1-\varepsilon_n)\log n},
\end{equation}
with $\varepsilon_n\to 0$ as $n\to 0$, so that $\bar{c}_n$ serves as a uniform upper bound
to $c\equiv c(\by,\bz)$ for every $(\by,\bz)\in\calY\times\calZ^n$, according to
\cite[eq.\ (6)]{ZL78}. 
Thus,
\begin{eqnarray}
P[\calE_{\mbox{\tiny t}}(\by,\bz)]&=&\sum_{\by'\in \calE_{\mbox{\tiny t}}(\by,\bz)}P(\by')\\
&\ge&\sum_{\by'\in \calE_1(\by,\bz)}P(\by')\\
&\ge&\sum_{\by'\in \calE_1(\by,\bz)}P(\by',\tilde{\bs})\\
&=&\sum_{\by'\in \calE_1(\by,\bz)}P(\by,\tilde{\bs})\\
&=&|\calE_1(\by,\bz)|\cdot P(\by,\tilde{\bs})\\
&\ge&K^{-c}\cdot |E_1(\by,\bz)|\cdot P(\by)\\
&\ge&K^{-\bar{c}_n}\cdot |E_1(\by,\bz)|\cdot P(\by).
\end{eqnarray}
Now, let $\calT(\by|\bz,\hat{\bt},\tilde{\bs})$ 
denote the set of all $\by'\in\calY^n$ that are obtained
from $\by$ by permuting $\by$--phrases, $\{y_{n_i+1}^{n_{i+1}}\}$,
that are: (i) aligned to the same $\bz$-phrases, $z_{n_i+1}^{n_{i+1}}$, 
(ii) of the same length, (iii) begin at the same states,
of both $\hat{t}_i=(\hat{\theta}_{n_{i}},\hat{\sigma}_{n_{i}},\hat{\omega}_{n_{i}})$ and
$\tilde{s}_i=(\tilde{\theta}_{n_{i}},\tilde{\omega}_{n_{i}})$, 
and (iv) end at the same states of both
$\hat{t}_{i+1}=(\hat{\theta}_{n_{i+1}},\hat{\sigma}_{n_{i+1}},\hat{\omega}_{n_{i+1}})$ and
$\tilde{s}_{i+1}=(\tilde{\theta}_{n_{i+1}},\tilde{\omega}_{n_{i+1}})$.
Clearly, $\calT(\by|\bz,\hat{\bt},\tilde{\bt})\subseteq \calE_1(\by,\bz)$, and so,
$P[\calE_{\mbox{\tiny t}}(\by,\bz)]$ is further lower bounded by
\begin{equation}
P[\calE_{\mbox{\tiny t}}(\by,\bz)]\ge K^{-\bar{c}_n}
|\calT(\by|\bz,\hat{\bt},\tilde{\bt})|\cdot
P(\by).
\end{equation}
Now, according to Lemma 1 of \cite{Ziv85},
\begin{equation}
|\calT(\by|\bz,\hat{\bt},\tilde{\bt})|\ge
\exp_2\{v(\by,\bz)-n\epsilon_2^\prime(n)\},
\end{equation}
where 
\begin{eqnarray}
\epsilon_2^\prime(n)&=&
\frac{\bar{c}_n}{n}\cdot\log(|\Theta|^4\cdot|\Omega|^4\cdot|\Sigma|^2e)\\
&=&
\frac{\log(|\calY|\cdot|\calZ|)}{(1-\varepsilon_n)\log n}\cdot
\log(|\Theta|^4\cdot|\Omega|^4\cdot|\Sigma|^2e)\\
&=&O\left(\frac{1}{\log n}\right).
\end{eqnarray}
Thus,
\begin{equation}
P[\calE_{\mbox{\tiny t}}(\by,\bz)]\ge K^{-\bar{c}_n}
P(\by)\cdot \exp_2\{v(\by,\bz)-n\epsilon_2^\prime(n)\}
\dfn \exp_2\{u(\by,\bz)-n\epsilon_2(n)\}
\end{equation}
where 
\begin{eqnarray}
\epsilon_2(n)&=&\epsilon_2^\prime(n)+\frac{\bar{c}_n\log K}{n}\\
&\le&
\epsilon_2^\prime(n)+\frac{\log(|\calY|\cdot|\calZ|)\cdot
\log K}{(1-\varepsilon_n)\log n}\\
&=&O\left(\frac{1}{\log
n}\right),
\end{eqnarray}
and so,
\begin{equation}
\label{pet}
P[\calE_{\mbox{\tiny t}}(\by,\bz)]\ge
\exp_2\{u(\by,\bz)-n\epsilon_2(n)\}.
\end{equation}
To complete the proof, we use the first part of Lemma 1 of \cite{LaZ98}, which
asserts that for every $a,b\in[0,1]$, $f(a)/f(b)\le\max\{1,a/b\}$, and so,
\begin{eqnarray}
\frac{f(P[\calE_{\mbox{\tiny u}}(\by,\bz)])}
{f(P[\calE_{\mbox{\tiny t}}(\by,\bz)])}&\le&
\max\left\{1,\frac{P[\calE_{\mbox{\tiny u}}(\by,\bz)]}{P[\calE_{\mbox{\tiny
t}}(\by,\bz)]}\right\}\\
&\le&\max\left\{1,\frac{\exp_2\{u(\by,\bz)+n\epsilon_1(n)\}}
{\exp_2\{u(\by,\bz)-n\epsilon_2(n)\}}\right\}\\
&\le&e^{n[\epsilon_1(n)+\epsilon_2(n)]},
\end{eqnarray}
where in the second inequality, we have used eqs.\ (\ref{peu}) and (\ref{pet}).
Now, referring to Lemma \ref{threshold}, let us define
\begin{eqnarray}
\epsilon_3(n)&=&\frac{1}{n}\log\left\{\left(\frac{K}{\pi_{\min}}\right)^{2\bar{c}_n}
\left[n\ln\left(\frac{1}{\pi_{\min}|\Theta\times\Sigma|}\right)+1\right]+1\right\}\\
&=&O\left(\frac{1}{\log n}\right).
\end{eqnarray}
Then,
\begin{eqnarray}
\bar{\mbox{P}}_{\mbox{\tiny e,o}}(\bz)&\ge&
e^{-n\epsilon_3(n)}
\bar{\mbox{P}}_{\mbox{\tiny e,t}}(\bz)~~~~~~
~~~~~~\mbox{(by Lemma
\ref{threshold})}\\
&=&e^{-n\epsilon_3(n)}
\sum_{\by\in\calY^n}P(\by|\bz)f(P[\calE_{\mbox{\tiny t}}(\by,\bz)])\\
&\ge&e^{-n[\epsilon_1(n)+\epsilon_2(n)+\epsilon_3(n)]}
\sum_{\by\in\calY^n}P(\by|\bz)f(P[\calE_{\mbox{\tiny u}}(\by,\bz)])\\
&=&e^{-n[\epsilon_1(n)+\epsilon_2(n)+\epsilon_3(n)]}\bar{\mbox{P}}_{\mbox{\tiny
e,u}}(\bz).
\end{eqnarray}
Finally, upon averaging both sides 
over $\{\bz\}$, we complete the proof of Theorem 1, with
\begin{equation}
\epsilon(n)\dfn\epsilon_1(n)+\epsilon_2(n)+\epsilon_3(n),
\end{equation}
which is $O((\log\log n)/\log n)$ since 
$\epsilon_1(n)$ is such.

\section*{Appendix}
\renewcommand{\theequation}{A.\arabic{equation}}
    \setcounter{equation}{0}

\subsection*{A1. Proof of Lemma \ref{threshold}}

Let us define
\begin{eqnarray}
\Delta(\by,\bz)
&\dfn&\{\by':~M_{\mbox{\tiny o}}(\by',\bz)> M_{\mbox{\tiny o}}(\by,\bz),
~P(\bz|\by')\ge \alpha^{-1}P(\bz|\by)\}\\
&=&\{\by':~M_{\mbox{\tiny o}}(\by',\bz)> M_{\mbox{\tiny o}}(\by,\bz),
~P(\by)P(\by'|\bz)\ge \alpha^{-1}P(\by')P(\by|\bz)\},
\end{eqnarray}
so that $\calE_{\mbox{\tiny t}}(\by,\bz)$ is given by the disjoint union of
$\calE_{\mbox{\tiny o}}(\by,\bz)$ and $\Delta(\by,\bz)$.
Then the average conditional error probabilities given $\bz$ are
\begin{eqnarray}
\bar{P}_{\mbox{\tiny e,o}}(\bz)&=&\sum_{\by}P(\by|\bz)f(P[\calE_{\mbox{\tiny
o}}(\by,\bz)])\\
\bar{P}_{\mbox{\tiny
e,t}}(\bz)&=&\sum_{\by}P(\by|\bz)f(P[\calE_{\mbox{\tiny o}}(\by,\bz)]+P[\Delta(\by,\bz)])\\
&\le&\sum_{\by}P(\by|\bz)\left(\frac{P[\calE_{\mbox{\tiny o}}(\by,\bz)]+
P[\Delta(\by,\bz)]}{(P[\calE_{\mbox{\tiny
o}}(\by,\bz)]}\right)f(P[\calE_{\mbox{\tiny o}}(\by,\bz)]),
\end{eqnarray}
where in the last step, we have used the first part of Lemma 1 from \cite{LaZ98}
(see also \cite{FL98}).
Now, let us define
\begin{equation}
r(\by,\bz)\dfn\sum_{\by'\in\calE_{\mbox{\tiny o}}(\by,\bz)}P(\by'|\bz).
\end{equation}
Then,
\begin{eqnarray}
P(\by)&=&\sum_{\by'}P(\by)P(\by'|\bz)\\
&\ge&\sum_{\by'\in\calE_{\mbox{\tiny o}}(\by,\bz)}P(\by)P(\by'|\bz)+
\sum_{\by'\in\Delta(\by,\bz)}P(\by)P(\by'|\bz)\\
&=&P(\by)r(\by,\bz)+
\sum_{\by'\in\Delta(\by,\bz)}P(\by)P(\by'|\bz)\\
&\ge&P(\by)r(\by,\bz)+\frac{1}{\alpha}
\sum_{\by'\in\Delta(\by,\bz)}P(\by')P(\by|\bz)\\
&=&P(\by)r(\by,\bz)+\frac{P(\by|\bz)}{\alpha}
P[\Delta(\by,\bz)],
\end{eqnarray}
and so,
\begin{equation}
P(\by|\bz)P[\Delta(\by,\bz)]\le \alpha P(\by)[1-r(\by,\bz)].
\end{equation}
We then have
\begin{eqnarray}
& &\bar{P}_{\mbox{\tiny e,t}}(\bz)-\bar{P}_{\mbox{\tiny e,o}}(\bz)\\
&\le&\sum_{\by}P(\by|\bz)\frac{P[\Delta(\by,\bz)]}
{P[\calE_{\mbox{\tiny o}}(\by,\bz)]}f(P[\calE_{\mbox{\tiny o}}(\by,\bz)])\\
&\le&\alpha\cdot\sum_{\by}\frac{P(\by)[1-r(\by,\bz)]}
{P[\calE_{\mbox{\tiny o}}(\by,\bz)]}f(P[\calE_{\mbox{\tiny o}}(\by,\bz)])\\
&=&\alpha\cdot\sum_{\by}\sum_{\{\by':~M_{\mbox{\tiny o}}(\by',\bz) >
M_{\mbox{\tiny o}}(\by,\bz)\}}\frac{P(\by)P(\by'|\bz)}
{P[\calE_{\mbox{\tiny o}}(\by,\bz)]}f(P[\calE_{\mbox{\tiny o}}(\by,\bz)])\\
&\eqa&\alpha\cdot\sum_{\by'}\sum_{\{\by:~M_{\mbox{\tiny o}}(\by',\bz) > M_{\mbox{\tiny
o}}(\by,\bz)\}}
\frac{P(\by)P(\by'|\bz)}{P[\calE_{\mbox{\tiny
o}}(\by,\bz)]}f(P[\calE_{\mbox{\tiny o}}(\by,\bz)])\\
&\leb&\alpha\cdot\sum_{\by'}\sum_{\{\by:~M_{\mbox{\tiny o}}(\by',\bz) > M_{\mbox{\tiny
o}}(\by,\bz)\}}
\frac{P(\by)P(\by'|\bz)}{P[\calE_{\mbox{\tiny o}}(\by,\bz)]}
f(P[\calE_{\mbox{\tiny o}}(\by',\bz)])\\
&\le&\alpha\cdot\sum_{\by'}P(\by'|\bz)f(P[\calE_{\mbox{\tiny o}}(\by',\bz)])\cdot\sum_{\by}
\frac{P(\by)}{P[\calE_{\mbox{\tiny o}}(\by,\bz)]}\\
&=&\alpha\cdot\bar{P}_{\mbox{\tiny
e,o}}(\bz)\cdot\sum_{\by\in\calY^n}\frac{P(\by)}{P[\calE_{\mbox{\tiny o}}(\by,\bz)]},
\end{eqnarray}
where in (a) we have interchanged the order of the summation and in (b), we
have used the monotonicity of $f$ together with the fact that
$\calE_{\mbox{\tiny o}}(\by,\bz)\subseteq \calE_{\mbox{\tiny o}}(\by',\bz)$
whenever $M_{\mbox{\tiny o}}(\by',\bz) > M_{\mbox{\tiny o}}(\by,\bz)$.
To complete the proof, it remains to show then that
for any $\bz$,
\begin{equation}
L_n(\bz)\dfn \sum_{\by\in\calY^n}\frac{P(\by)}{P[\calE(\by,\bz)]}=
\sum_{\by\in\calY^n}\frac{P(\by)}{\sum_{\{\by':~M_{\mbox{\tiny o}}(\by',\bz) \le
M_{\mbox{\tiny o}}(\by,\bz)\}}P(\by')}
\end{equation}
cannot exceed $1+n\ln[1/(\pi_{\min}|\Theta\times\Omega|)]$.
For the given $\bz$, consider the ordering of all members of $\calY^n$
(not only those in $\calC$)
according to the ranking function $M_{\mbox{\tiny o}}(\by,\bz)$, i.e.,
\begin{equation}
P(\bz|\by[1])\ge
P(\bz|\by[2])\ge\ldots\ge P(\bz|\by[N]),~~~~~~~~N=|\calY|^n
\end{equation}
and let us denote $a_i=P(\by[i])$, $A_i=\sum_{j=1}^i a_j$, $i=1,\ldots,N$.
Then, using the facts that $A_1=a_1=P(\by[1])$ and $A_N=1$, as well as the inequality
\begin{equation}
\ln(1+u)\equiv-\ln\left(1-\frac{u}{1+u}\right)\ge\frac{u}{1+u},
\end{equation}
we have
\begin{eqnarray}
L_n(\bz)&=&\sum_{i=1}^N\frac{a_i}{A_i}\\
&=&1+\sum_{i=2}^N\frac{a_i}{A_{i-1}+a_i}\\
&=&1+\sum_{i=2}^N\frac{a_i/A_{i-1}}{1+a_i/A_{i-1}}\\
&\le&1+\sum_{i=2}^N\ln\left(1+\frac{a_i}{A_{i-1}}\right)\\
&=&1+\sum_{i=2}^N\ln\left(\frac{A_{i-1}+a_i}{A_{i-1}}\right)\\
&=&1+\sum_{i=2}^N\ln\left(\frac{A_i}{A_{i-1}}\right)\\
&=&1+\ln\left(\frac{A_N}{A_1}\right)\\
&=&\ln\left[\frac{1}{P(\by[1])}\right]+1\\
&\le&\ln\left[\frac{1}{(\pi_{\min} \cdot|\Theta|\cdot|\Omega|)^n}\right]+1\\
&=&n\ln\left(\frac{1}{\pi_{\min}\cdot|\Theta|\cdot|\Omega|}\right)+1,
\end{eqnarray}
where we have used the assumption (\ref{positivity}),
which implies that
$P(\by)\ge(\pi_{\min}\cdot|\Theta|\cdot|\Omega|)^n$ for all $\by$.
This completes the proof of Lemma \ref{threshold}.

\subsection*{A.2 Proof of Eq.\ (\ref{zm92})}

We next show that for every $\by$ and $\bs$,
\begin{equation}
P(\by)\le
P(\by,\bs)\cdot\left(\frac{|\Theta\times\Omega|}{\pi_{\min}^2}\right)^c.
\end{equation}
For the sake of brevity, let us denote $\zeta_i=(\theta_i,\omega_i)$
(so that $s_i=\zeta_{n_i}$).
Now,
\begin{equation}
P(\by,\bs)=\prod_{i=0}^{c-1}
\pi(y_{n_i+1}^{n_{i+1}},\zeta_{n_{i+1}}|\zeta_{n_i}).
\end{equation}
But
\begin{eqnarray}
\pi(y_{n_i+1}^{n_{i+1}},\zeta_{n_{i+1}}|\zeta_{n_i})
&=&\sum_{\zeta_{n_i+1}^{n_{i+1}-1}}\prod_{t=n_i+1}^{n_{i+1}}\pi(y_t,\zeta_t|\zeta_{t-1})\\
&=&\sum_{\zeta_{n_i+1}}\pi(y_{n_i+1},\zeta_{n_i+1}|\zeta_{n_i})\times\nonumber\\
& &\sum_{\zeta_{n_i+2}^{n_{i+1}-2}}
\prod_{t=n_i+2}^{n_{i+1}-1}\pi(y_t,\zeta_t|\zeta_{t-1})\times\nonumber\\
& &\sum_{\zeta_{n_{i+1}-1}}\pi(y_{n_{i+1}},\zeta_{n_{i+1}}|\zeta_{n_{i+1}-1})\\
&\ge&\pi_{\min}^2\sum_{\zeta_{n_i+1}^{n_{i+1}-1}}
\prod_{t=n_i+2}^{n_{i+1}-1}\pi(y_t,\zeta_t|\zeta_{t-1}),
\end{eqnarray}
where we have assumed that $n_i+2 \le n_{i+1}-1$, which means that the
phrase length must be at least three,\footnote{This assumption does not affect the
generality, as the number of phrases of length shorter than three
cannot exceed $|\calY\times\calZ|+|\calY\times\calZ|^2$, which is fixed and
hence negligible compared to the total number of phrases for large $n$.}
and where we have lower bounded both $\pi(y_{n_i+1},\zeta_{n_i+1}|\zeta_{n_i})$ and 
$\pi(y_{n_{i+1}},\zeta_{n_{i+1}}|\zeta_{n_{i+1}-1})$ by $\pi_{\min}$.
Similarly, since both $\pi(y_{n_i+1},\zeta_{n_i+1}|\zeta_{n_i})$ and 
$\pi(y_{n_{i+1}},\zeta_{n_{i+1}}|\zeta_{n_{i+1}-1})$ are upper bounded by
unity, we have
\begin{equation}
\pi(y_{n_i+1}^{n_{i+1}},\zeta_{n_{i+1}}|\zeta_{n_i})\le
\sum_{\zeta_{n_i+1}^{n_{i+1}-1}}
\prod_{t=n_i+2}^{n_{i+1}-1}\pi(y_t,\zeta_t|\zeta_{t-1}).
\end{equation}
Since the expression
$$\sum_{\zeta_{n_i+1}^{n_{i+1}-1}}
\prod_{t=n_i+2}^{n_{i+1}-1}\pi(y_t,\zeta_t|\zeta_{t-1})$$
depends neither on $\zeta_{n_i}$ nor on $\zeta_{n_{i+1}}$, it follows
that for any $\zeta_{n_i}$,
$\zeta_{n_i}^\prime$,
$\zeta_{n_{i+1}}$, and
$\zeta_{n_{i+1}}^\prime$,
\begin{equation}
\pi_{\min}^2\le\frac{\pi(y_{n_i+1}^{n_{i+1}},\zeta_{n_{i+1}}^\prime|\zeta_{n_i}^\prime)}
{\pi(y_{n_i+1}^{n_{i+1}},\zeta_{n_{i+1}}|\zeta_{n_i})}\le
\frac{1}{\pi_{\min}^2},
\end{equation}
and so,
\begin{eqnarray}
P(\by)&=&\sum_{\bs^\prime}P(\by,\bs^\prime)\\
&=&P(\by,\bs)\sum_{\bs^\prime}\frac{P(\by,\bs^\prime)}{P(\by,\bs)}\\
&=&P(\by,\bs)\sum_{\bs^\prime}\prod_{i=0}^{c-1}
\frac{\pi(y_{n_i+1}^{n_{i+1}},\zeta_{n_{i+1}}^\prime|\zeta_{n_i}^\prime)}
{\pi(y_{n_i+1}^{n_{i+1}},\zeta_{n_{i+1}}|\zeta_{n_i})}\\
&\le&P(\by,\bs)\sum_{\bs^\prime}\prod_{i=0}^{c-1}\frac{1}{\pi_{\min}^2}\\
&=&P(\by,\bs)\cdot\left(\frac{|\Omega\times\Theta|}{\pi_{\min}^2}\right)^c,
\end{eqnarray}
which completes the proof of eq.\ (\ref{zm92}).


\end{document}